# Asymmetric electron distribution induced intrinsically strong anisotropy of thermal transport in bulk CrOCl

Qikun Tian,[1] Qi Yang,[1] An Huang,[1] Bo Peng,[1] Jinbo Zhang,[1] Xiong Zheng,[1] Jian Zhou,[1] Zhenzhen Qin,[2] and Guangzhao Qin[1*]

[1] *State Key Laboratory of Advanced Design and Manufacturing Technology for Vehicle, College of Mechanical and Vehicle Engineering, Hunan University, Changsha 410082, P. R. China*

[2] *Key Laboratory of Materials Physics, Ministry of Education, School of Physics, Zhengzhou University, Zhengzhou 450001, P. R. China*

[*] gzqin@hnu.edu.cn

# Abstract

Anisotropic heat transfer offers promising solutions to the efficient heat dissipation in the realm of electronic device thermal management. However, the fundamental origin of the anisotropy of thermal transport remains mysterious. In this paper, by combining frequency domain thermoreflectance (FDTR) technique and first-principles based multiscale simulations, we report the intrinsic anisotropy of thermal transport in bulk CrOCl, and further trace the origin of the anisotropy back to the fundamental electronic structures. The in-plane and cross-plane thermal conductivities ($\kappa$) at 300 K are found to be 21.6 and 2.18 Wm$^{-1}$K$^{-1}$, respectively, showcasing a strong $\kappa_{in\text{-}plane}/\kappa_{cross\text{-}plane}$ ratio of ~10. Deep analysis of orbital-resolved electronic structures reveals that electrons are mainly distributed along the in-plane direction with limited interlayer distribution along the cross-plane direction, fundamentally leading to the intrinsic anisotropy of thermal transport in bulk CrOCl. The insight gained in this work would shed light on the design of advanced thermal functional materials.

# INTRODUCTION

The escalating power density and operating frequency of power electronic equipment have recently triggered a commensurate surge in the thermal output of semiconductor devices during high-power operations, thereby accentuating the prominence of heat dissipation challenges [1–5]. To ensure the consistent and reliable performance of semiconductor device chips, heat dissipation is required to be managed skillfully and effectively. Notably, anisotropic materials have superior heat transfer capabilities to eliminate hotspots along the fast-axis (with high thermal conductivity $\kappa$) and provide thermal insulation along the slow-axis (with low $\kappa$) [6,7], which has the potential to facilitate directional heat transfer and significantly enhance the thermal management and overall thermal dissipation efficiency. Consequently, the exploration and development of materials characterized by anisotropic heat transfer capabilities emerge as a promising approach for addressing thermal challenges and the practical applications in heat dissipation and advanced thermal management of electronic devices. However, although there has been some research on thermal transport anisotropy, it lacks systematic and in-depth exploration. Moreover, the fundamental origin of the anisotropy of thermal transport remains mysterious.

To date, the common techniques for measuring anisotropic thermal conductivity based on electrothermal and optical pump probe principles include the 3-omega method [8,9], micro-Raman thermometry [10], microfabricated suspended devices [11], frequency-domain thermoreflectance (FDTR) [12–18], time-domain thermoreflectance (TDTR) [19–21], *etc*. Among these measurement methods, the FDTR possesses the advantages of low costs and simplicity of operation without the requirement of complicated microfabrication compared to 3-omega, micro-Raman thermometry, and microfabricated suspended device methods, and no need for ultrafast pulsed laser or a delay stage in comparison to TDTR. Moreover, FDTR method has the ability to simultaneously characterize the in-plane and cross-plane thermal conductivities by fitting the thermal model at a wide range of modulation frequencies. Thus, it is of significant advantages to employ FDTR technique to characterize anisotropic thermal conductivity of materials.

Additionally, the recently emerged first-principles calculations offer a profound fundamental analysis for understanding and forecasting the thermal transport properties of materials from the fundamental level of atomic and electronic structures [22–24]. When integrated with finite element models (FEMs), it further enables multiscale research spanning from the microscopic to the macroscopic level. Compared to molecular dynamics (MD) simulations using empirical potentials to model interatomic interactions [25–27] and experiments with the requirement of equipment and surroundings, first-principles based multiscale simulations offer both accuracy and cost-effectiveness, serving as a powerful tool for the study of thermal transport properties.

The layered *van der Waals* (vdW) materials are of great interest in the field of anisotropic heat transfer and thermal management [7,19,28–35], such as graphite [19,28] and few-layer graphene [33], transition metal dichalcogenide (TMD) $MoX_2$ [29,30] and $WX_2$ [30–32] (X = S, Se, Te), black phosphorus (BP) [34] and phosphorene [35], *etc*. The sublayers in vdW materials are connected by vdW forces, and the weak interlayer vdW interactions usually result in the lower thermal conductivity along the cross-plane direction compared to the in-plane direction [29]. As one of the representative vdW materials, the bulk CrOCl with antiferromagnetic (AFM) properties has garnered much attention due to the potential applications in spintronics, optoelectronics, and nanoelectronics [36–40]. The anisotropic layered vdW structure makes it a competitive candidate for the anisotropic heat transfer applications. Very recently, Guo *et al.* fabricated the $MoS_2$/CrOCl heterostructures to reconfigure the carrier polarity in $MoS_2$ from n- to p-type via strong vdW interfacial coupling, enabling the construction of vertically integrated complementary logic circuits with stable performance [41]. Since effective thermal management is essential to ensure device stability and prolonged lifetime in high-performance nanoelectronic devices, understanding the anisotropic thermal transport properties of CrOCl is of great significance to optimize the thermal management and improve the performance of heterojunction devices. However, up to date, the systematic investigation on the anisotropic thermal transport properties of bulk CrOCl is still lacking, which hinders its practical applications. Particularly, previous research on the anisotropic thermal conductivity of materials has predominantly employed either experimental measurements [28,30,31] or theoretical calculations [22–24], with relatively limited exploration into combining both methods to investigate anisotropic thermal transport properties and fundamentally uncover the underlying mechanisms. Therefore, exploring the intrinsic anisotropic thermal conductivity of bulk CrOCl and its underlying physical mechanisms by combining both experimental measurement and theoretical calculation is on urgent demand, which is vital for the advanced thermal management in nanoelectronic devices.

In this study, by utilizing the FDTR experimental technique combined with the first-principles based multiscale simulations, we explore the intrinsic anisotropic thermal transport properties of bulk CrOCl. The in-plane and cross-plane thermal conductivities of bulk CrOCl at 300 K are measured by FDTR, which are further verified by first-principles calculations. To reveal the underlying physical mechanisms and the origin of anisotropy in thermal transport, the mechanical properties, phonon transport properties, and electronic structures of bulk CrOCl are further examined. It is found that the asymmetric electron distribution fundamentally induces intrinsic anisotropy of thermal transport in bulk CrOCl. Finally, the FEMs are established to simulate the anisotropic thermal transport at device level. Our study provides insights for a deep understanding of the anisotropic thermal transport properties of bulk CrOCl, which holds significant promise for practical applications in directional heat transfer and advanced thermal management.

## RESULTS AND DISCUSSION

The vdW bulk CrOCl with the *Pmmn* space group possesses an orthorhombic structure and exhibits AFM semiconductor properties [39,40,42]. Figure 1(a) demonstrates that the primitive cell of bulk CrOCl consists of two Cr, two O, and two Cl atoms, where each Cr atom is bonded by neighboring four O and two Cl atoms. The Cr-O double sublayers are sandwiched between two Cl atom sublayers. Moreover, the primitive cell features the 2D rectangle sublattice in the *a–b* plane and a large vdW gap along the *c*-axis. The optimized lattice constants are *a* = 3.256 Å, b = 3.940 Å, and c = 7.820 Å, which are in good accordance with previous results [39,42]. Upon analyzing the photograph and scanning electron microscopy (SEM) images of the bulk CrOCl single crystal sample across various scales [Figure 1(b) and 1(c)], it is evident that the surface exhibits a remarkably smooth morphology at the microscale. This feature provides a solid foundation and ensures the reliability of the FDTR measurements. For further analyzing the chemical composition of the sample, the energy-dispersive X-ray spectroscopy (EDS) was used to obtain elemental maps of Cr, O, and Cl, as exhibited in Figure 1(d)-(g). It can be observed that these elements are uniformly distributed in the sample, conforming to the chemical stoichiometric ratio. As shown in Figure 1(h), the Raman spectrum of bulk CrOCl contains three main $A_g$ modes, which are located in 206.5 ($A_g^1$), 413.3 ($A_g^2$), and 457.6 cm$^{-1}$ ($A_g^3$), agreeing well with previous reports [36,38].

To confirm the dynamical and thermal stabilities of bulk CrOCl, the phonon spectrum was calculated and *ab initio* molecular dynamics (AIMD) simulations at 300 K were also conducted. The results as shown in Figures 1(i) and 1(j) indicate that no imaginary frequency is observed, and the free energy fluctuates within a narrow scope. It is also found that the equilibrium structure is found without significant distortion or bond-breaking, implying that the bulk CrOCl is dynamically and thermally stable. Moreover, the optical phonon branch frequencies with Raman activity are highlighted by the gray region in phonon dispersions, which correspond essentially to the three peaks of Raman spectrum [Figure 1(d)]. Note that the acoustic phonon branches of TA (transverse acoustic branch) and LA (longitudinal acoustic branch) along the Γ-X, Γ-Y, and Γ-Z directions (in accord with the *x*, *y*, *z* directions in real space) exhibit distinct slope with the steepest along Γ-X direction, then Γ-Y direction, and the smallest along Γ-Z direction. Due to the distinct slopes of the phonon dispersion along different directions arising from the vdW feature in lattice, the phonon group velocities will be also different, promising that the bulk CrOCl is an anisotropic thermal conductivity material.

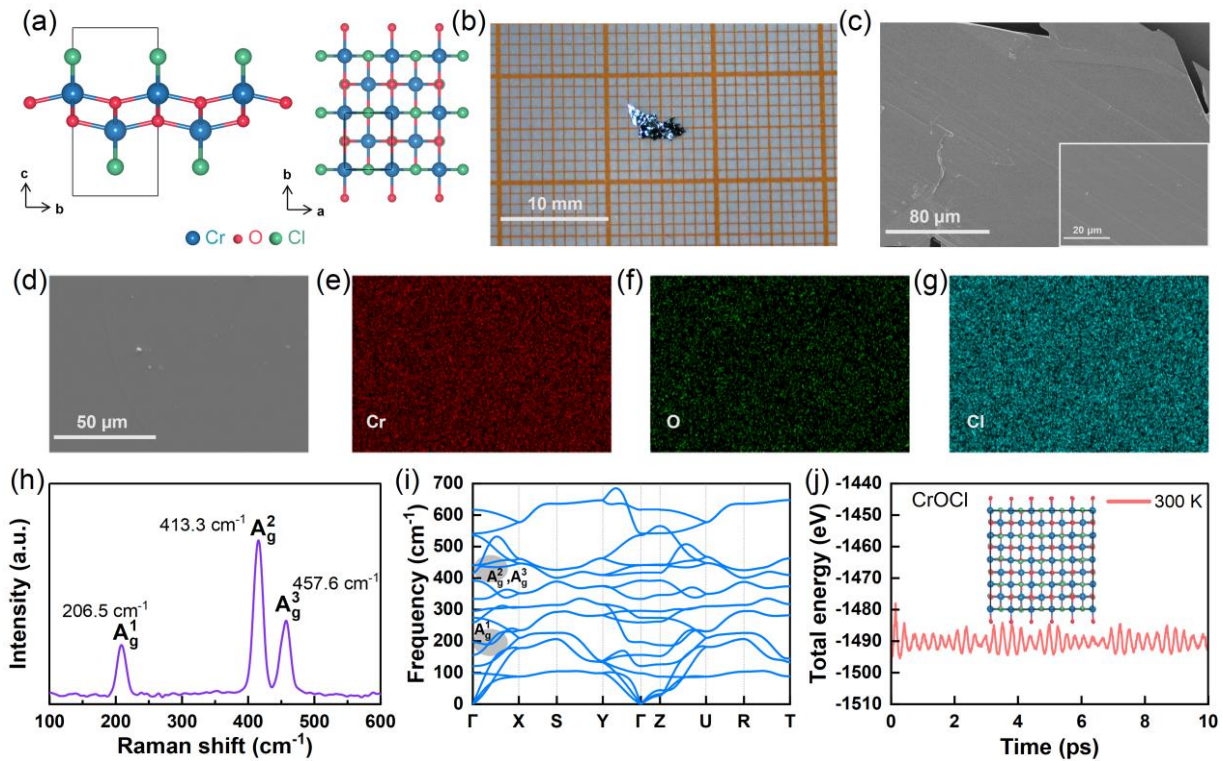

**Figure 1** (a) The side and top views of the crystal structure of bulk CrOCl. The blue, red, and green balls denote Cr, O, and Cl atoms, respectively. (b) Photograph of CrOCl sample. (c) SEM images of the bulk CrOCl single crystal sample. (d) The SEM image of the analyzed area and the corresponding elements' distribution of the (e) Cr, (f) O, and (g) Cl in the related area by EDS. (h) The Raman spectrum of bulk CrOCl. (i) The calculated phonon spectrum of bulk CrOCl, where the optical phonon branches with Raman activity are highlighted by the gray region. (j) The AIMD simulations of total energy fluctuation in bulk CrOCl lasting for 10 ps at 300 K, where the inset is the lattice structure after the equilibrium simulations.

The heat transfer properties of materials are significant for the application in the fields of advanced thermal management [1–4] and thermoelectrics [5]. To study the thermal transport properties of bulk CrOCl, we utilized the FDTR technique to experimentally measure its thermal conductivity, and deep analysis are achieved by first-principles based multiscale simulations.

The FDTR technique was employed to characterize the thermal transport properties of materials using two continuous wave laser beams as pump beam and probe beam [12–17], respectively. Figures 2(a, b) display that the schematic of FDTR setup and beam offset measurement of the focused probe spot's radius. For accurate measurements, the sample is coated with a thin Au layer (~100 nm) as the transducer layer. The Au layer efficiently absorbs a considerable percentage of the pump laser's intensity, causing a gradual temperature increase and the formation of a temperature gradient spreading to the underlying sample [17]. The temperature change of the sample surface induces a corresponding change in its optical reflectivity, resulting in variations in the modulated intensity of the continuous-wave probe laser reflected from the layer surface. The lock-in amplifier is used to record the amplitude and phase of the reflected optical signal at multiple pump beam modulation frequencies. Moreover, the phase lag between the pump and probe beam is also measured, where the pump signal is regarded as reference signal. To acquire the thermal transport properties of sample such as the $\kappa$ and thermal boundary conductance (TBC), the layered heat diffusion model [18] is employed with the requirement of materials parameters such as the heat capacity and layer thickness.

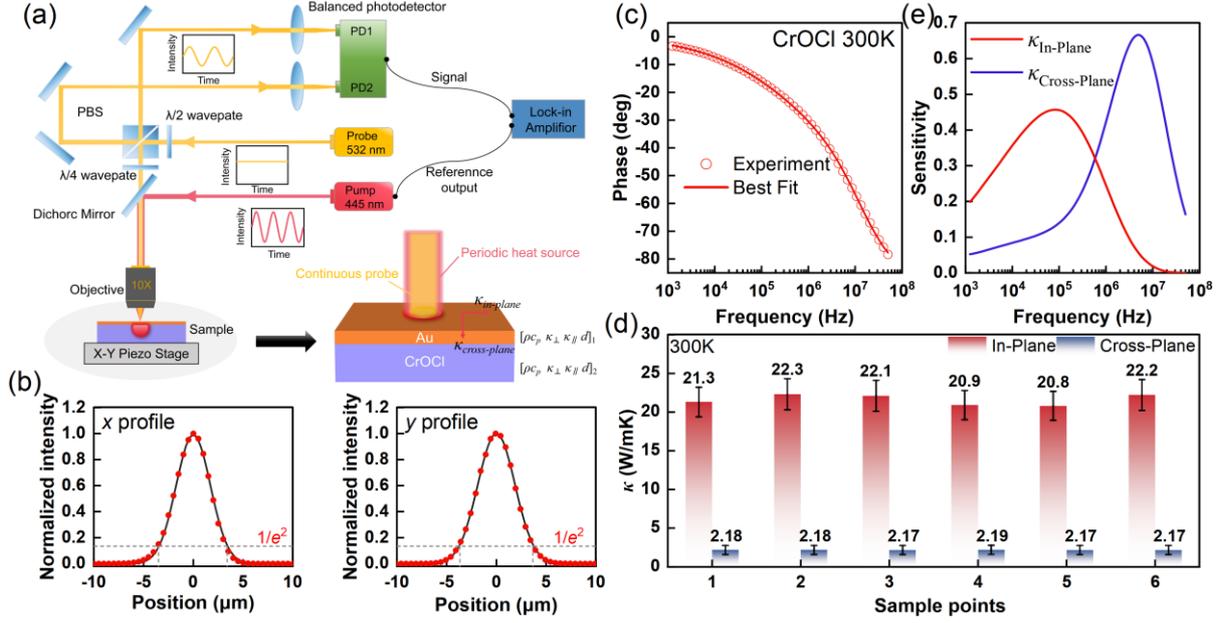

**Figure 2** (a) Schematic of FDTR setup and sample configuration. (b) Beam offset measurement of the focused probe spot's radius in both *x* and *y* directions. (c) FDTR phase data with respect to frequency from 1 kHz to 50 MHz for 100 nm Au coated CrOCl sample at 300 K. (d) In-plane and cross-plane thermal conductivities by fitting the FDTR thermal model for six independent sample points. (e) Sensitivity of thermal phase to thermal transport properties as a function of frequency. The spot radius used for sensitivity analysis is ~3.6 μm based on the beam offset measurements.

Figure 2(c) shows the measured phase data with respect to modulation frequency from 1 kHz to 50 MHz for bulk CrOCl sample. The thickness, heat capacity, and thermal conductivity of Au and the heat capacity of bulk CrOCl are derived by experimental measurements and theoretical calculations. Thus, the remaining parameters to be determined are the $\kappa$ of bulk CrOCl and the TBC of Au/CrOCl interface, with a focus on the in-plane and cross-plane $\kappa$ of bulk CrOCl. To ensure the accuracy of the measured thermal conductivity, we performed six independent measurements at random points on the bulk CrOCl sample. Figure S1 shows that the phase signals for the six independent measurements and Figure 2(d) displays the results of in-plane and cross-plane thermal conductivity fitted by the FDTR thermal model for six independent sample points. The average cross-plane $\kappa$ is measured as 2.18 ± 0.01 $Wm^{-1}K^{-1}$ and in-plane $\kappa$ is 21.6 ± 0.68 $Wm^{-1}K^{-1}$ for bulk CrOCl. It is found that the $\kappa$ of bulk CrOCl along the in-plane direction is much larger than that along the cross-plane direction, which shows a strong anisotropy and the anisotropy value of $\kappa_{in\text{-}plane}/\kappa_{cross\text{-}plane}$ is about 10. Moreover, sensitivity analysis is carried out to determine the sensitivity of the measured phase signals to the various parameters of Au/CrOCl, including the in-plane $\kappa$, cross-plane $\kappa$, and TBC, as plotted in Figure 2(e) and Figure S2. The results show that the modulation frequency has a significant impact on the recorded phase signal's sensitivity. The cross-plane $\kappa$ is mainly extracted from the model fit at a high frequency range (1 to 50 MHz), while the in-plane $\kappa$ is extracted at a low frequency range (1 kHz to 1 MHz). Our measured signals cover a frequency range that includes both in-plane and cross-plane thermal conductivity sensitive intervals, ensuring the reliability of measurement results. In short, based on the FDTR measurements, the bulk CrOCl is an anisotropic thermal conductivity material with promising applications in directional heat dissipation and advanced thermal management of electronic devices.

To further study the anisotropic thermal conductivity of bulk CrOCl and uncover the origin of the anisotropic thermal transport properties, first-principles calculations are performed and the $\kappa$ of bulk CrOCl is obtained through solving the phonon Boltzmann transport equation (BTE) using ShengBTE code [43]. Since the main heat carriers in semiconductors are phonons and the contribution of electrons to the thermal conductivity is negligible, we mainly focus on the lattice thermal conductivity in the following discussions. The calculated thermal conductivities of bulk CrOCl are compared to the experimental results. Moreover, deep analysis is performed to achieve fundamental understanding on the anisotropic thermal transport properties.

To ensure the calculation accuracy of lattice thermal conductivity of bulk CrOCl, the convergence of $\kappa$ is fully tested on the basis of cutoff distance, nearest neighbor and $Q$-grid, as shown in Figure 3. The interatomic interactions include the harmonic interatomic force constants (second-IFCs) and high-order IFCs (third-IFCs and higher). A long enough cutoff distance is key to obtaining converged $\kappa$. However, it is huge resource consuming to acquire the high-order IFCs compared to the harmonic IFCs. Thus, the root-mean-square (RMS) for harmonic IFCs is utilized to evaluate the cutoff distance based on the estimation of the distance dependent strength of interatomic interactions. The RMS is expressed of elements as [44,45]:

$$\text{RMS}(\Phi_{ij}) = \left[\frac{1}{9}\sum_{\alpha,\beta}\left(\Phi_{ij}^{\alpha\beta}\right)^2\right]^{\frac{1}{2}}, \qquad (1)$$

where the $\Phi_{ij}^{\alpha\beta}$ is the second-order IFCs tensor, representing the harmonic response of the force acting on atom $i$ ($\alpha$-direction) induced by the influence of atom $j$ ($\beta$-direction) displacement. A large RMS($\Phi_{ij}$) value means the strong anharmonic interactions. Figure 3(a) shows that the RMS($\Phi_{ij}$) of bulk CrOCl decreases with the cutoff distance increasing, which converges when the distance reaches 4 Å. As shown in Figure 3(b), the $\kappa$ decreases with the increasing nearest neighbors due to the fact that more interatomic interactions and phonon–phonon scattering are considered, where the convergence trend agrees very well with the distance dependent strength of interatomic interactions [Figure 3(a)]. Moreover, we test the $Q$-grid to ensure the $\kappa$ tends to remain stable, as shown in Figure 3(c). With a full convergence test on the basis of nearest neighbors and $Q$-grid, the results exhibit that the 7$^{th}$ nearest neighbor corresponding to the cutoff distance with 4.18 Å is enough to describe the anharmonic interatomic interactions and acquire the converged $\kappa$.

The $\kappa$ of bulk CrOCl at 300 K is predicted to be 23.2 and 2.31 Wm$^{-1}$K$^{-1}$ along the in-plane and cross-plane directions, respectively, agreeing well with the experimental measure results, as shown in Figure 3(d). The calculated results demonstrate that the experimentally measured thermal conductivity of CrOCl using the FDTR is reliable. Additionally, the temperature-dependent thermal conductivity change curves are shown in Figure 3(e), which basically agree with the $T^{-1}$ decreasing law. The results reveal that the phonon Umklapp scattering is dominant in the phonon scattering processes. Besides, the $\kappa$ of CrOCl exhibits strong anisotropy along the in-plane and the cross-plane directions with the anisotropic ratio being ~10. It is observed that the anisotropic ratio $\kappa_{in\text{-}plane}/\kappa_{cross\text{-}plane}$ slightly decreases with decreasing temperature as plotted in Figure 3(f). The decreased anisotropy at low temperatures could be due to the different cutoff frequencies of the LA branches along in-plane and cross-plane orientations, resulting in variable phonon scattering and then influencing thermal conductivity. Specifically, along the in-plane direction, because of the higher cutoff frequency, the peak of phonon thermal conductivity occurs more rapidly as the temperature decreases. As approaching the peak, the increase of thermal conductivity slows down. Consequently, the anisotropy of thermal conductivity decreases with the decreasing temperature.

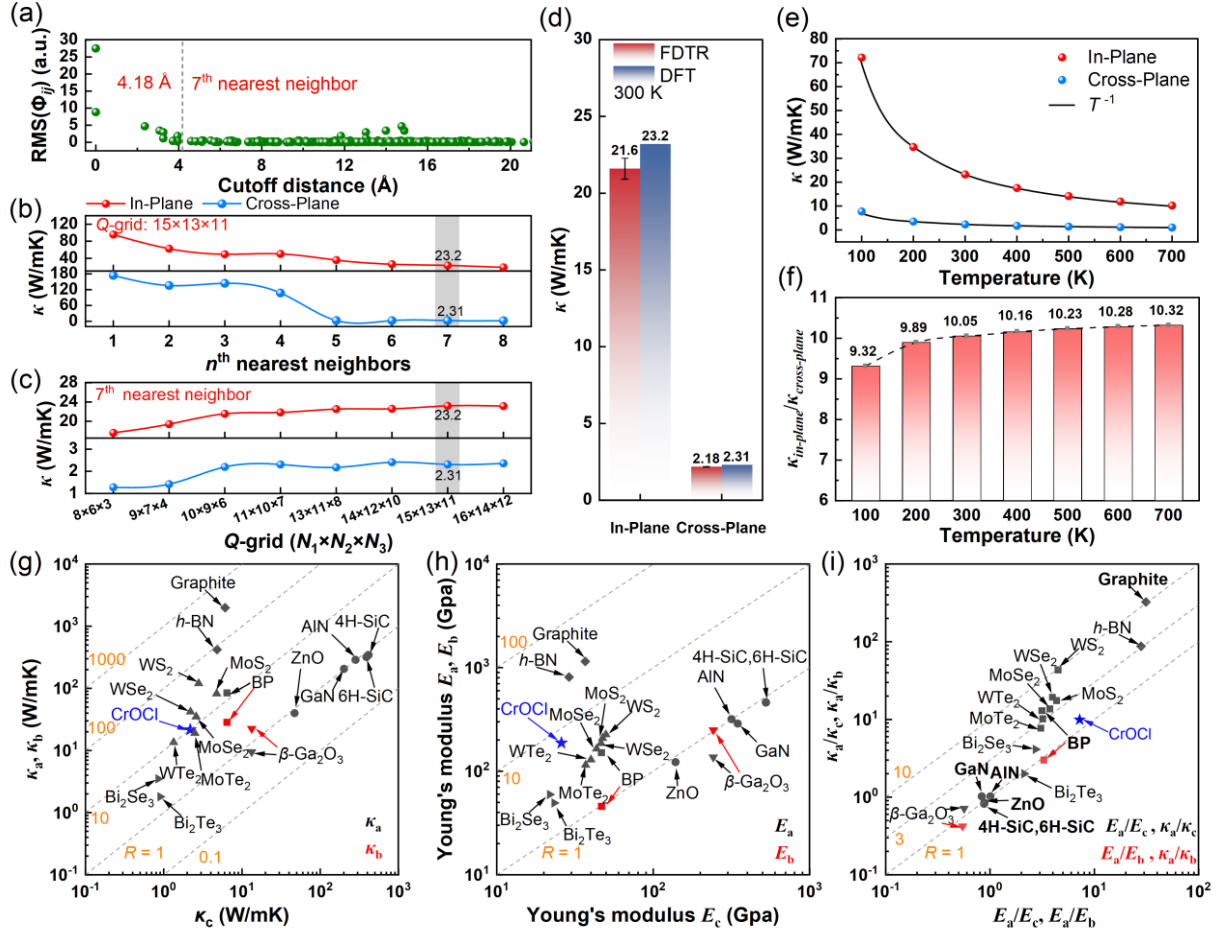

**Figure 3** (a) The RMS with respect to the cutoff distance. (b) Convergence of the thermal conductivity with the nearest neighbor. (c) Convergence of the thermal conductivity with the *Q*-grid. The grey areas mark the converged value. (d) In-plane and cross-plane thermal conductivity of bulk CrOCl at 300 K from FDTR measurements and DFT calculations. Temperature-dependent (e) thermal conductivities and (f) anisotropic ratio $\kappa_{in\text{-}plane}/\kappa_{cross\text{-}plane}$. (g) Measured intrinsic anisotropic thermal conductivities of typical bulk materials [19,29,30,32,34,46–52] via various experimental methods. (h) Calculated anisotropic Young's modulus of typical bulk materials [24,53–56]. (i) The relationship between the anisotropy of Young's modulus and thermal conductivity. The dashed lines refer to the constant ratio of anisotropy of R (dividing the y-coordinate by the x-coordinate).

To further study the anisotropy of thermal transport in bulk CrOCl, we collect in Figure 3(g) the anisotropic thermal conductivities of bulk vdW materials such as graphite [19], BP [34], *h*-BN [46], transition metal dichalcogenides (TMDs) MX$_2$ (M = Mo, W; X=S, Se, Te) [29,30,32], Bi$_2$Se$_3$ [47], Bi$_2$Te$_3$ [48], CrOCl, and non-vdW bulk materials including 4H-SiC [49], 6H-SiC [49], AlN [50], GaN [50], ZnO [51], *β*-Ga$_2$O$_3$ [52]. The anisotropy ratio of CrOCl is comparable to that of widely focused TMDs family. Compared to non-vdW bulk materials, bulk vdW materials usually possess higher anisotropy ratios of thermal conductivity, as also listed in Table S1. The difference in anisotropy are originated from the fact that the layers of the bulk vdW materials are connected by weak vdW forces, while the atoms in non-vdW bulk are all covalently bonded. Among these systems, graphite has the highest anisotropy ratio up to ~300, with in-plane and cross-plane thermal conductivities being ~2000 and ~6 Wm$^{-1}$K$^{-1}$, respectively [19]. Note that BP [34] and $\beta$-Ga$_2$O$_3$ [52] exhibit not only cross-plane anisotropy ($\kappa_a/\kappa_c$) but also in-plane anisotropy ($\kappa_a/\kappa_b$).

The elastic constants and Young's modulus can quantify the bonding strength, which are key factors influencing the thermal transport in essence. Thus, the mechanical properties including elastic constants and anisotropic Young's modulus of all the above bulk materials are calculated, as summarized in Figure 3(h) and Table S2. It can be seen in Figures 3(g, h, i) that the anisotropy exhibited by Young's modulus overall agrees well with the anisotropy exhibited by thermal conductivity. Especially, the Young's modulus of bulk CrOCl along in-plane ($E_a$, $E_b$) and cross-plane directions ($E_c$) are 187.5 and 25.9 GPa, respectively, where the anisotropy of Young's modulus is well consistent with the anisotropy of thermal conductivity.

Figure 3(i) further elucidates the relationship between the anisotropy of Young's modulus and thermal conductivity. Specifically, graphite demonstrates strongest thermal conductivity anisotropy, which correlates well with its significant Young's modulus anisotropy (The constant ratio of anisotropy R ≈ 10, where R=anisotropy of thermal conductivity/anisotropy of Young's modulus). This can be attributed to the fact that the interlayer atoms in graphite are connected by the typical vdW forces along the vertical direction. In contrast, the non-vdW bulk materials such as AlN, GaN, ZnO, 4H-SiC, and 6H-SiC exhibit weak thermal conductivity anisotropy, corresponding to their minimal Young's modulus anisotropy (R ≈ 1), which is originated from the fact that atoms are connected by covalent bonds. The law as discussed above is well-verified by the anisotropy of BP along the in-plane and cross-plane directions, where the anisotropy's magnitude is directly correlated with the way of atom connections. It can be seen that the in-plane anisotropic behavior of BP is similar to the non-vdW bulk materials (R ≈ 1), which is due to the covalent bonding of in-plane atoms. Different from the in-plane anisotropic behavior, the cross-plane anisotropic behavior of BP is consistent with the typical vdW bulk materials (R ≈ 3 ~ 10) due to the vdW forces connection, such as TMDs family MX$_2$ and $h$-BN. Notably, the anisotropy of thermal conductivity in CrOCl is comparable to that in the TMDs family, while CrOCl exhibits smaller R lying between the typical vdW and non-vdW bulk materials. The characteristic lies in the interlayers of CrOCl are connected by the strongly polarized vdW forces of two Cl atoms and the strength is between covalent bonds and vdW forces. Based on the comprehensive discussion presented above, the anisotropic thermal conductivity properties of bulk CrOCl is well understood.

To comprehensively understand the underlying physical mechanism of anisotropic thermal transport in bulk CrOCl, we further conduct calculations and discussions on the phonon transport properties. It is well known that the lattice thermal conductivity along the $\alpha$-direction ($\alpha = x, y, z$) can be acquired by summing contribution from all the phonon modes ($\lambda$) [43]:

$$\kappa_\alpha = \frac{1}{V}\sum_\lambda C_\lambda v_{\lambda\alpha}^2 \tau_{\lambda\alpha},\qquad(2)$$

where $V$ is the lattice volume of the unit cell, and $C_\lambda$, $v_{\lambda\alpha}$, and $\tau_{\lambda\alpha}$ represent the specific heat capacity, group velocity and relaxation time, respectively. Since the phonon relaxation time and specific heat have no direction dependence [57], the anisotropy of thermal conductivity can be attributed to the different phonon group velocities in distinct directions. Figure 4(a) displays that the phonon group velocity along the in-plane direction is clearly larger than phonon group velocity along the cross-plane direction at 300 K. The anisotropy behavior of phonon group velocity agrees well with the slopes of phonon dispersions as shown in Figure 1(i), which is consistent with the anisotropy behavior of $\kappa$ as shown in Figure 3(d). Furthermore, the phonon mean free path (MFP) is plotted in Figure 4(b). It is found that the $\kappa$ values converge to that of the infinite system when the MFP reaches about $10^4$ nm for both the in-plane and cross-plane directions. Although the upper limit for the phonon MFP along the cross-plane direction is slightly larger than the in-plane direction, the relatively lower phonon group velocity along the cross-plane direction leads to the corresponding smaller $\kappa$. Moreover, the phonon relaxation time is inversely related to phonon scattering rate, depending on both the scattering strength and scattering probability, which can be evaluated by the Grüneisen parameter and scattering phase space, respectively. Figure 4(d) displays that the Grüneisen parameter of bulk CrOCl at low frequencies is large, implying the strong phonon anharmonicity. The phase space is related to the accessible scattering channels for the three-phonon scattering process, where the larger phase space means more available scattering channels. As plotted in Figure 4(e), the phase space of bulk CrOCl reaches up to $10^{-7}$ at low-frequency range. The integration of large Grüneisen parameter and phase space leads to the small phonon relaxation time as shown in Figure 4(f).

To uncover the origin of $\kappa$ anisotropy from the view of phonon branches, the percentage contribution to $\kappa$ from different branches are extracted as shown in Figure 4(c). It is clearly shown that the $\kappa$ along the cross-plane direction is mainly contributed by the acoustic phonon branches, which agrees well with most semiconductors such as silicon, GaN, *etc*. However, for the in-plane direction, it is unusual to find that the optical phonon branches anomalously dominate $\kappa$. To uncover the origin, we focus on the thermal transport properties in the high-frequency region where optical phonon branches locate. It is clearly shown that the in-plane optical phonon branches at high frequencies has a higher group velocity [Figure 4(a)]. Meanwhile, the corresponding high-frequency optical phonon branches has a small scattering phase space [Figure 4(e)]. Thus, the combination of large group velocity and small scattering phase space leads to the large contributions to the in-plane thermal conductivity from the optical phonon branches.

To achieve a comprehensive understanding of the intrinsic anisotropy in thermal transport observed in bulk CrOCl, it is essential to adopt an analytical approach that extends beyond phonon transport. This approach thoroughly explores the electronic structures, as these ultimately govern the material's properties. The orbital projected band structure and projected density of states (*p*DOS) are calculated to further analyze the physical mechanism and origin of the anisotropic properties at the level of electronic structure, as shown in Figure 4(g). The spin-up and spin-down bands are identical due to the AFM feature of bulk CrOCl. The calculated bandgap is 2.63 eV, which is consistent with previous reports [39,40]. Moreover, it is observed that the conduction and valence band dispersion along the Γ-Z path (corresponding to *z*-direction in the real space) is much flatter than that along the Γ-X path (corresponding to *x*-direction in the real space) and Γ-Y path (corresponding to *y*-direction in the real space), indicating the highly anisotropic electronic and crystal structure properties. Note that the conduction band minimum (CBM) is mainly contributed by the Cr_*d* orbitals, and the valence band maximum (VBM) is dominated by the Cl_*p* and O_*p* orbitals.

The electron localization functions (ELF) of bulk CrOCl in the top and side views is shown in Figure 4(h). It is revealed that the distribution of electrons is predominantly oriented along the in-plane direction, with a notably limited interlayer distribution along the cross-plane direction, implying the anisotropic electronic structure properties. Particularly, the Bader analysis [58] is further employed to quantitatively investigate the charge distribution and transfer in bulk CrOCl. The charge transfer for Cr-O and Cr-Cl covalent bonds are 1.19 *e* and 0.54 *e*, respectively, where the O and Cl atoms gain electrons and Cr atoms lose electrons due to the O atoms have the strongest electronegativity. The total number of electrons lost by the Cr atom with a value of 1.73 *e* is equal to the sum of the electrons gained by the O and Cl atoms, exhibiting that the electrons in bulk CrOCl transfer within sublayers and no electron transfer between the sublayers along the cross-plane direction, as shown in Figure 4(h,i). Thus, the anisotropic phonon thermal transport properties are well understood from the asymmetric electron distribution in bulk CrOCl.

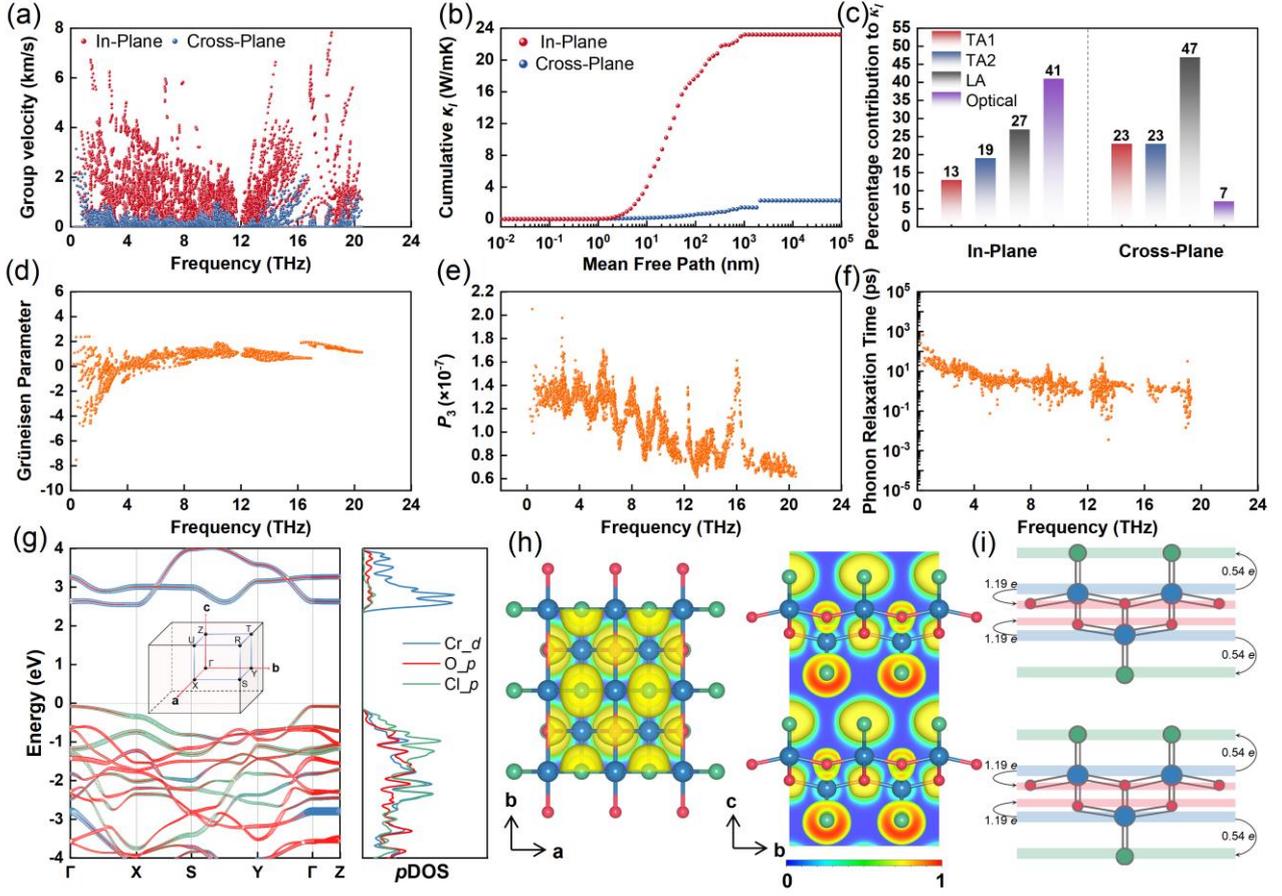

**Figure 4** The anisotropic thermal transport properties of bulk CrOCl at 300 K. (a) phonon group velocity. (b) cumulative lattice thermal conductivity with respect to the phonon MFP. (c) The percentage contribution to $\kappa$ from different branches including the acoustic phonon modes (TA1, TA2, LA) and optical phonon modes. The (d) Grüneisen parameter, (e) $P_3$ phase space, and (f) phonon relaxation time as the function of frequency. (g) Orbital projected band structures and $p$DOS employing DFT+$U$ method. Considering the AFM properties of bulk CrOCl, the spin-up and spin-down bands are same and $p$DOS is displayed with only the spin-up part. The inset shows the first Brillouin zone. (h) Top and side views of the ELF. (i) Schematic diagram of charge transfer in bulk CrOCl.

To elucidate the practical applications of anisotropic heat transfer in device cooling, we further conduct FEM simulations and analyses of the directional heat transfer process at device level based on bulk CrOCl. Figure 5(a) displays the schematic of anisotropic heat transfer in bulk CrOCl based device. The overall initial temperature of established FEMs is 120 °C. By setting up a cold end with a temperature of 25 °C in the in-plane and cross-plane directions, respectively, a temperature gradient for heat flow transfer is created, as shown in Figure 5(b). The results demonstrate that the heat flows faster along the in-plane direction than that along the cross-plane direction. Figure 5(c) shows the temperature variation in the in-plane and cross-plane directions as a function of time. Consistent with the anisotropic heat flow, the temperature drops more quickly along the in-plane direction than that along the cross-plane direction. The fundamental reason lies in that the in-plane direction has a higher thermal conductivity than the cross-plane direction, which is in line with our computational and experimental results of bulk CrOCl. Thus, bulk CrOCl with strong anisotropic thermal conductivity enables directional heat transfer to rapidly eliminate hotspots, which is promising for the applications in the field of heat dissipation and advanced thermal management of nanoelectronic devices.

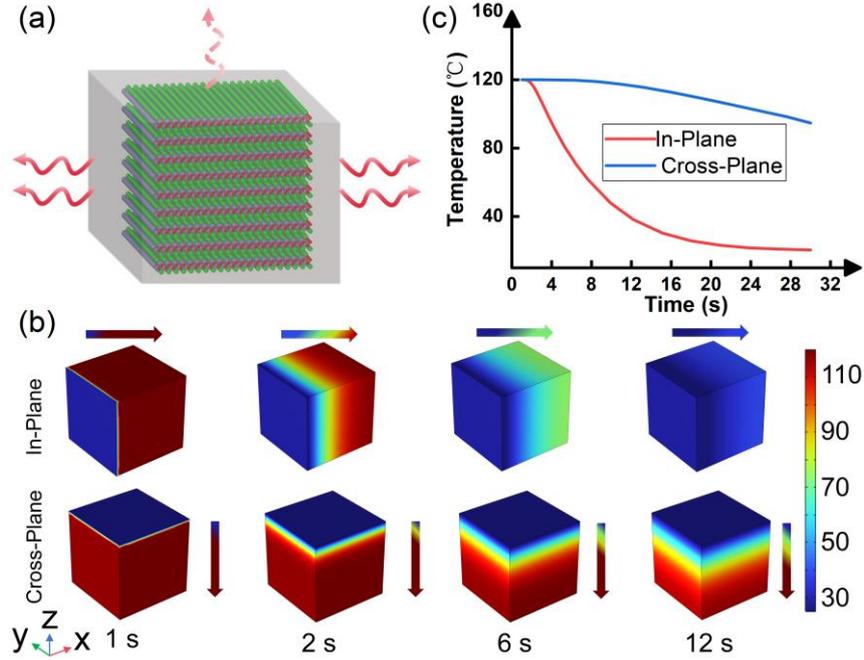

**Figure 5** (a) Schematic of anisotropic heat transfer in bulk CrOCl based device. The dashed arrow represents slow heat transfer, while solid arrows represent fast heat transfer. (b) Schematic of anisotropic thermal transport by FEM simulations. (c) Temperature evolution with respect to time along the in-plane and cross-plane directions, respectively.

## CONCLUSION

In summary, utilizing FDTR combined with first-principles based multiscale simulations, we systematically explored the intrinsic anisotropic thermal transport properties of bulk CrOCl. The in-plane and cross-plane thermal conductivities of bulk CrOCl at 300 K are measured as 21.6 and 2.18 Wm$^{-1}$K$^{-1}$, respectively, with the strong anisotropic ratio of $\kappa_{in\text{-}plane}/\kappa_{cross\text{-}plane}$ about 10. To uncover the fundamental physical mechanism for the anisotropic $\kappa$ of bulk CrOCl, first-principles based multiscale simulations are conducted. With a full convergence test, the $\kappa$ of bulk CrOCl at 300 K is calculated as 23.2 and 2.31 Wm$^{-1}$K$^{-1}$ along in-plane and cross-plane directions, respectively, agreeing well with the measured $\kappa$. Based on the comprehensive comparison with lots of anisotropic materials, the relationship between the anisotropy of Young's modulus and thermal conductivity is uncovered. The anisotropy of thermal conductivity in CrOCl is comparable to that in the TMDs family, while CrOCl exhibits smaller ratio of R= (anisotropy of thermal conductivity)/(anisotropy of Young's modulus) lying between the typical vdW and non-vdW bulk materials. The characteristic lies in the interlayers of CrOCl are connected by the strongly polarized vdW forces of two Cl atoms and the strength is between covalent bonds and vdW forces. Moreover, according to the analysis of phonon transport properties, it is found that the larger $\kappa$ along the in-plane direction in bulk CrOCl is dominated by the higher phonon group velocities compared to the cross-plane direction with lower phonon group velocities. Furthermore, the orbital band structures, ELF, and charge transfer are calculated to essentially understand the intrinsic anisotropy at the fundamental level of electronic structures. It is found that the distribution of electrons is mainly along the in-plane direction with limited interlayer distribution along the cross-plane direction. Thus, the intrinsic anisotropy of thermal transport in bulk CrOCl is fundamentally induced by the asymmetric electron distribution. Finally, FEMs are established to study the practical applications of anisotropic heat transfer in device cooling by simulating the directional heat transfer process at device level based on bulk CrOCl. The heat flow is transferred faster along the in-plane direction than the cross-plane direction, leading to the temperature dropping more quickly along the in-plane direction. The insight gained in this work would shed light on the design of advanced thermal functional materials. And the bulk CrOCl with strong anisotropic $\kappa$ is promising for the practical applications in directional heat transfer and advanced thermal management.

## METHODS

**Sample characterization and FDTR measurement technique**

The bulk CrOCl sample was purchased from the Shanghai Onway Technology Company, which was synthesized through the chemical vapor transport (CVT) method. We verified the composition of the samples by WITEC Raman microscope Alpha 300R [Figure 1(h)], where the observed three peaks and their intensities were consistent with previous reports [36,38]. The EDS was further employed to analyze the compositional distribution of various elements [Figures 1(d)-(g)]. The HITACHI SU8220

SEM instrument was used to characterize the microscopic morphology of the sample surface.

The FDTR technique was utilized to measure the in-plane and cross-plane thermal conductivities of bulk CrOCl. A 445 nm pump laser with incident power (~30 mW) was used to heat up the sample surface with transducer layer, resulting in the periodically changed surface temperature. A 532 nm probe laser with incident power (~5 mW) was utilized to detect the optical reflectance owing to the changed surface temperature. To efficiently absorb pump heat and match with the 532 nm wavelength of the probe laser, a 100-nm-thick gold layer was deposited on the CrOCl sample by electron beam evaporation. The phase of thermoreflectance signal was obtained by the lock-in amplifier and balanced photodetector, which was recorded as a function of modulation frequency and fitted to the heat conduction model [18]. The spot radius of laser, the thickness and thermal conductivity of the gold layer, and the heat capacity of gold layer and substrate were demanded as input parameters in the heat conduction model. Through beam offset measurements, the spot radius of laser was 3.6 μm ($1/e^2$ intensity level) along both $x$ and $y$ directions. The thermal conductivity of the gold layer was set as 180 Wm$^{-1}$K$^{-1}$ [59]. The heat capacity of $2\times10^6$ Jm$^{-3}$K$^{-1}$ of bulk CrOCl was calculated by first-principles calculations. With the pre-obtained parameters, the thermal conductivity of bulk CrOCl can be evaluated by fitting the measured thermoreflectance signals.

**First-principles based multiscale simulations**

First-principles calculations were carried out using the Vienna *ab initio* simulation package (VASP) based on the density functional theory (DFT) [60,61]. The electron exchange-correlation effect was described by the generalized gradient approximation (GGA) [62] of the Perdew-Burke-Ernzerhof (PBE) functional [63,64] and projected-augmented wave (PAW) method [65]. The plane-wave energy cutoff was chosen as 500 eV. The convergence thresholds of total energy and force were set to be $10^{-6}$ eV and $10^{-3}$ eV/Å, respectively. The DFT+$U$ approach with the value of $U-J$ = 5 eV was employed to handle the strong correlation effect of Cr_3$d$ orbitals using the Dudarev's approach [66]. For both ion relaxation and static calculations, a 16 × 13 × 6 Gamma-point centered grid was used to sample the Brillouin zone. Moreover, the DFT-D3 method was considered in all calculations for the vdW correction [67,68]. Furthermore, we verified the thermal stability of the bulk CrOCl using a supercell of 5×4×2 via AIMD simulations with a canonical ensemble (NVT) at 300 K for 10 ps [69]. The lattice thermal transport properties were studied by solving the phonon Boltzmann transport equation using Phonopy package [70] and ShengBTE code [43], where the 5×4×2 supercell containing 240 atoms were used for calculating the second and third-order IFCs. The convergence of both nearest neighbor and $Q$-grid was fully tested to achieve accurate lattice thermal conductivity. The charge transfer was analyzed by the Bader techniques [58].

The FEMs are constructed using COMSOL Multiphysics 6.2 (the heat transfer in solids interface module) to simulate the anisotropic heat transfer in bulk CrOCl based device. The 3D dimensions of the model were 10 mm×10 mm×10mm, and the overall initial temperature was 120 °C. The in-plane and cross-plane thermal conductivities obtained from first-principles calculations are adopted as input parameters in the constructed models. By setting up one end of the box face with 25 °C in the in-plane and cross-plane directions, respectively, a temperature gradient for heat flow transfer is established. Probes were set on the surface of the high temperature end to record the temperature variation.

## ACKNOWLEDGEMENTS

This work is supported by the National Natural Science Foundation of China (Grant No. 52006057), the National Key R&D Program of China (2023YFB2408100), the Fundamental Research Funds for the Central Universities (Grant No. 531119200237), the State Key Laboratory of Advanced Design and Manufacturing for Vehicle Body at Hunan University (Grant No. 52175013), the Natural Science Foundation of Chongqing, China (No. CSTB2022NSCQ-MSX0332), Outstanding Youth Project (23B0024) of Hunan Provincial Department of Education. Z.Q. is supported by the National Natural Science Foundation of China (Grant No.12274374). The numerical calculations in this paper have been done on the supercomputing system of the E.T. Cluster and the National Supercomputing Center in Changsha and Zhengzhou. We would like to thank Ms. Ying Liu (College of Mechanical and Vehicle Engineering, Hunan University, Changsha, China) for the help with Raman spectrum measurement.

## AUTHOR CONTRIBUTIONS

G.Q. supervised the project; Q.T. carried out the work and led the manuscript writing; Q.Y., A.H., B.P. and J. Zhang contributed to the discussion and data analysis; X.Z., J. Zhou, H.D. and Z.Q. provided insightful suggestions and reviewed the manuscript; and all the authors contributed to the final revision of this manuscript.

## COMPETING INTERESTS

The authors declare no competing interests.